\documentclass[aps,pre,twocolumn,superscriptaddress]{revtex4-2}
\usepackage{physics,bm}
\usepackage{graphicx}

\begin{document}


\title{Phase Transitions in a Modified Ising Spin Glass Model: \\
A Tensor-Network-based Sampling Approach}

\author{Takumi Oshima}
\affiliation{
 Graduate School of Arts and Sciences,
 The University of Tokyo,
 Komaba, Meguro-ku, Tokyo 153-8902, Japan
}
\author{Yamato Arai}
\affiliation{
 Graduate School of Arts and Sciences,
 The University of Tokyo,
 Komaba, Meguro-ku, Tokyo 153-8902, Japan
}
\author{Koji Hukushima}
\affiliation{
 Graduate School of Arts and Sciences,
 The University of Tokyo,
 Komaba, Meguro-ku, Tokyo 153-8902, Japan
}
\affiliation{
Komaba Institute for Science, The
University of Tokyo, 3-8-1 Komaba, Meguro-ku, Tokyo 153-8902, Japan
}

\date{\today}

\begin{abstract}
The possible reentrant ferromagnetic phase boundary in the two-dimensional Edwards-Anderson (EA) model raises the question of whether an intermediate Mattis-like spin-glass (SG) phase emerges in the Kitatani model, a special case of a modified Nishimori model with correlated disorder. In this model, generating bond configurations is computationally demanding because of the correlated random interactions. We address this difficulty by combining an exact gauge-based hierarchical representation of the bond distribution with independent tensor-network-based sampling of conditional spin configurations. This approach allows high-precision calculations for square lattices of up to $256\times 256$ spins. The leading finite-size-scaling analyses of the SG and ferromagnetic Binder ratios yield slightly different estimates of the spin-glass and ferromagnetic transition points, with a difference exceeding their quoted statistical uncertainties. A direct examination of the system-size dependence of the squared order parameters, however, does not resolve a distinct regime with SG order but without ferromagnetic order. These results provide a quantitative characterization of the narrow transition region relevant to the proposed intermediate Mattis-like SG phase and a basis for testing the reentrant transition scenario.  
\end{abstract}


\maketitle

\section{Introduction}
\label{sec:introduction}
Spin glasses are widely recognized as model systems characterized by the coexistence of randomness and frustration. The Edwards-Anderson (EA) model~\cite{SFEdwards_1975}, together with its mean-field counterpart, the Sherrington-Kirkpatrick (SK) model~\cite{Sherrington-Kircpatrick1975}, has been studied extensively for decades. Consequently, its phases and phase transitions with respect to temperature and disorder strength have been clarified in significant detail. However, while mean-field models have been analyzed mathematically, many questions remain regarding the finite-dimensional EA model~\cite{SGbeyond, FarBeyond}. 

A prominent unsolved problem concerns the nature of the phase boundary between the ferromagnetic (FM) and non-ferromagnetic phases, such as the spin-glass (SG) and paramagnetic (PM) phases. Mean-field theory for the SK model, incorporating replica symmetry breaking, indicates that the phase boundary is vertical with respect to the disorder axis, meaning it remains parallel to the temperature axis. This finding is consistent with Nishimori's gauge theory based on local gauge transformations~\cite{Nishimori1981}, which rigorously proves for any dimension, including the mean-field limit, that this boundary must be either vertical or reentrant. In other words, the FM phase cannot bulge outward toward the disordered regime. 

\begin{figure}[b]
    \centering
    \includegraphics[width=\linewidth]{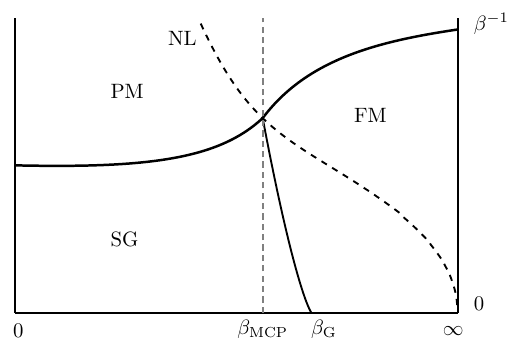}
    \caption{Generic schematic phase diagram for the Edwards-Anderson model. The vertical axis represents the temperature $T=\beta^{-1}$, and the horizontal axis indicates the disorder strength. The phase diagram shows paramagnetic (PM), ferromagnetic (FM), and spin-glass (SG) phases. The Nishimori line (NL), drawn as a dashed curve, passes through the multicritical point at $\beta_{\text{MCP}}$ on the phase boundary. The curve of the ferromagnetic phase boundary toward $\beta_\text{G}>\beta_{\text{MCP}}$ at low temperatures indicates the occurrence of a reentrant transition. The finite-temperature SG region shown here represents the generic situation in dimensions $d\geq 3$; in the two-dimensional EA model relevant to the present study, the conventional SG phase is confined to $T=0$.  }
    \label{fig:phase_diagram_EA}
\end{figure}

Furthermore, based on the Kitatani model~\cite{Kitatani1992}, a modified Ising spin-glass model, it has been argued under certain plausible assumptions that the phase boundary of the $\pm J$ Ising EA model should be strictly vertical. Conversely, numerical studies, including Markov-chain Monte Carlo (MCMC) simulations in two dimensions~\cite{ParisenToldin2009} and three dimensions~\cite{Giacomo2011}, real-space renormalization group analyses~\cite{Nobre2001}, and ground-state computations in two dimensions~\cite{Amoruso2004}, have consistently suggested the existence of a slight reentrant transition. The generic topology associated with such reentrant behavior is illustrated schematically in Fig.~\ref{fig:phase_diagram_EA}. The finite-temperature SG region shown in this schematic phase diagram is appropriate to dimensions $d\geq 3$; in the two-dimensional EA model, the SG phase is confined to the zero-temperature axis. 

While reentrant behavior has been observed in various frustrated spin models~\cite{Diep2013}, its physical origin and the conditions for its emergence in spin-glass systems remain under intense discussion. The reentrant phenomenon is particularly noteworthy because it contradicts the intuitive expectation that cooling increases order. Instead, it suggests a subtle competition where the entropy of disordered configurations plays a dominant role even at low temperatures. 

In a recent development, Nishimori proposed a modified spin-glass model~\cite{Nishimori2024} that establishes a direct connection between the reentrant transition and the temperature chaos at the limit of maximum disorder~\cite{Nishimorietal2025}. This model is particularly important because it generalizes the Kitatani model and suggests a scenario in which the Kitatani argument's underlying assumptions are violated. A distinguishing characteristic of this model is that the interaction bonds are inherently correlated. Unlike the conventional EA model, these bonds cannot be generated as independent and identically distributed (i.i.d.) variables from a simple probability distribution. Consequently, the generation of interaction bonds imposes a substantial computational barrier to numerical investigations. By re-examining the underlying gauge-theoretical structure, we demonstrate that these interaction sets can be generated by a specific sampling procedure. This allows both interactions and spin configurations to be treated as objects for sampling. 

Even with this framework, the efficiency of sampling remains a critical concern. Although the replica-exchange Monte Carlo method (parallel tempering)~\cite{Geyer1991, HukushimaNemoto, Marinari1998} is a standard tool for simulating spin glasses and can be adapted to the present problem, we instead employ a tensor-network-based sampling method that generates configurations independently. This sampling approach avoids the slow thermalization and autocorrelation issues frequently encountered in traditional MCMC methods. Using this analysis, we investigate the possible existence of an intermediate Mattis-like SG phase and its implications for the reentrant transition in the EA model.  

The rest of the paper is organized as follows. In Sec.~\ref{sec:model}, we define the modified Nishimori model, including the Kitatani model, and describe our numerical methodology based on the tensor-network hierarchical sampling scheme. Sec.~\ref{sec: Numerical_results} presents the Binder ratios and the second moments of order parameters through the finite-size-scaling analysis. We examine the possible separation between the onset of spin-glass and ferromagnetic order and estimate the corresponding critical exponents. Finally, Sec.~\ref{sec:conclusion} concludes the paper with a discussion on the physical implications of our results for the reentrant transition in two-dimensional spin glasses.

\section{Model and Methods}
\label{sec:model}
\subsection{Modified Nishimori spin-glass model}
We consider an Ising spin-glass model 
with $N=L^2$ spins on an $L\times L$ square lattice with free boundary conditions in both directions. The system is defined by the Hamiltonian 
\begin{equation}
    H(\bm{S}) = -\sum_{\langle ij\rangle}\tau_{ij}S_iS_j, 
\end{equation}
where $\bm{S}=(S_1,\ldots, S_N)$ denotes a spin configuration with $S_i=\pm 1$ at each site $i$, and $\tau_{ij}=\pm 1$ denotes the exchange interaction between sites $i$ and $j$. The summation runs over all the nearest-neighbor pairs. 

In the spin-glass model, two types of statistical averages are distinguished. For a fixed configuration of interactions $\bm{\tau}$, the thermal average of an observable $\mathcal{O}$ at inverse temperature $\beta$ is defined as 
\begin{equation}
    \langle \mathcal{O}\rangle_{\bm{\tau}} = \sum_{\bm{S}}\mathcal{O}(\bm{S}) \frac{e^{-\beta H(\bm{S})}}{Z_{\bm{\tau}}(\beta)},
\end{equation}
where $Z_{\bm{\tau}}(\beta)$ is the partition function:  
\begin{equation}
    Z_{\bm{\tau}}(\beta) = \sum_{\bm{S}}e^{-\beta H(\bm{S})}. 
\end{equation}

Unlike the standard EA model, where the interactions $\tau_{ij}$ are drawn independently from a fixed distribution, the modified Nishimori model~\cite{Nishimorietal2025} introduces inherent correlations between interaction bonds. The probability distribution of the bond configuration $\bm{\tau}$ is characterized by two parameters, $\gamma$ and $\beta_p$:  
\begin{equation}
    P(\bm{\tau}; \gamma,\beta_p) = \frac{Z_{\bm{\tau}}(\gamma)}{(2\cosh\gamma)^{N_B}}\frac{e^{\beta_p\sum_{\langle ij\rangle}\tau_{ij}}}{Z_{\bm{\tau}}(\beta_p)},  
    \label{eqn:distribution_tau}
\end{equation}
where $N_B$ is the total number of bonds. The presence of the partition function in the distribution induces non-trivial correlations between the bonds. 

The overall expectation value, which incorporates the disorder average with respect to the interactions $\bm{\tau}$, is denoted by $[\langle\mathcal{O}\rangle]$ and is obtained by averaging the thermal average $\langle\mathcal{O}\rangle_{\bm{\tau}}$ over the bond distribution $P(\bm{\tau}; \gamma,\beta_p)$: 
\begin{equation}
    [\langle \mathcal{O}\rangle ] = \sum_{\bm{\tau}}\langle \mathcal{O}\rangle_{\bm{\tau}}P(\bm{\tau}; \gamma,\beta_p). 
\end{equation}

When $\gamma=\beta_p$, the partition functions $Z_{\bm{\tau}}(\gamma)$ and $Z_{\bm{\tau}}(\beta_p)$ in the definition cancel each other out, and $P(\bm{\tau})$ reduces to the distribution of the standard EA model: 
\begin{equation}
    P(\bm{\tau};\beta_p,\beta_p) = \prod_{\langle ij\rangle}\frac{e^{\beta_p\tau_{ij}}}{2\cosh\beta_p}=P_{\rm EA}(\bm{\tau}; \beta_p).
    \label{eqn:distribution_EA}
\end{equation}
Here, $\beta_p$ denotes the Nishimori inverse temperature, which is related to the fraction of ferromagnetic bonds $p$ via $e^{2\beta_p} = {p}/{(1-p)}$. Meanwhile, the case $\gamma=0$ corresponds to the model introduced in Ref.~\cite{Nishimori2024}, where the distribution becomes
\begin{equation}
    P(\bm{\tau}; 0,\beta_p) \propto \frac{e^{\beta_p\sum_{\langle ij\rangle}\tau_{ij}}}{Z_{\bm{\tau}}(\beta_p)}.
\end{equation}

For $\gamma\neq \beta_p$, the distribution in Eq.~(\ref{eqn:distribution_tau}) involves a ratio of partition functions, implying that the bonds $\tau_{ij}$ are no longer independent. The partition function $Z_{\bm{\tau}}(\beta_p)$ in the denominator suppresses interaction bonds with large partition functions; instead, it enhances the statistical weight of bond configurations with high ground-state energies, thereby intensifying frustration. Conversely, $Z_{\bm{\tau}}(\gamma)$ in the numerator favors interaction bonds with weaker frustration. Thus, the parameter $\gamma$ plays a crucial role in controlling the strength of frustration, a distinctive feature absent in the standard EA model. 

\begin{figure}
    \centering
\includegraphics[width=\linewidth]{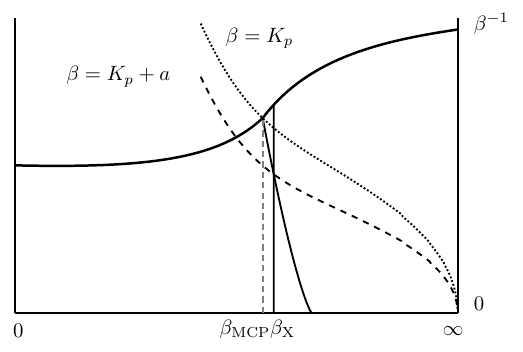}
    \caption{Generic schematic phase diagram illustrating the Kitatani model and its relation to the EA model. The vertical axis represents the temperature $T=\beta^{-1}$, and the horizontal axis indicates the disorder strength. The dashed curve denotes the Nishimori line for the Kitatani model, defined by $\beta=K_p+a$, while the dotted curve represents the Nishimori line for the corresponding EA model, $\beta=K_p$. The multicritical point $\beta_{\mathrm{MCP}}$ and the ferromagnetic transition point $\beta_\mathrm{X}$ are indicated on the dotted and dashed curves, respectively. 
    Under the reentrant-transition scenario, a possible intermediate Mattis-like spin glass phase would emerge along the dotted curve $\beta=K_p$, where the gauge and physical spin configurations are sampled at the different inverse temperatures $\beta_p=K_p+a$ and $\beta=K_p$, respectively, in the region $\beta_{\mathrm{MCP}}<\beta<\beta_{\mathrm{X}}$.  
    The diagram is intended to illustrate the general relation between the two models rather than a numerically determined phase diagram specific to two dimensions. In the standard two-dimensional EA model, the conventional SG phase is confined to the zero-temperature axis; the possible finite-temperature M-SG phase in the correlated-disorder Kitatani model is the subject here.  
}
    \label{fig:phase_diagram_Nishimori}
\end{figure}

The Kitatani model~\cite{Kitatani1992} is a specific case of this modified Nishimori model, corresponding to $\gamma=K_p$ and $\beta_p=K_p+a$:
\begin{equation}
    P(\bm{\tau}; K_p,K_p+a) =\frac{Z_\tau(K_p)}{(2\cosh K_p)^{N_B}}\frac{e^{(K_p+a)\sum_{\langle ij\rangle}\tau_{ij}}}{Z_\tau(K_p+a)}. 
\end{equation}
The schematic phase diagram of this Kitatani model is shown in Fig.~\ref{fig:phase_diagram_Nishimori}. 
Under the assumption that any ordered phase above the Nishimori line, $\beta=K_p+a$, in the Kitatani model is, if it exists, always ferromagnetic, it was argued that no reentrant transition should exist in the EA model~\cite{Kitatani1992}. However, this remains inconsistent with numerical simulations on the EA model that consistently suggest the presence of a reentrant transition~\cite{ParisenToldin2009, Giacomo2011}. 

If such a reentrant transition exists in the EA model, it would imply the emergence of an intermediate Mattis-like spin-glass (M-SG) phase in the modified model~\cite{Nishimorietal2025}. Here, an M-SG phase refers to a frozen phase characterized by a nonzero replica overlap but a vanishing uniform magnetization. In terms of the order parameters formally defined in Sec.~\ref{sec:physical_observables}, this corresponds to $q\neq 0$ and $m=0$. The terminology recalls the Mattis model~\cite{MATTIS1976421}, in which the random interactions can be written as $J_{ij}=J_0\xi_i\xi_j$ and transformed into uniform ferromagnetic interactions by a local gauge transformation. Although the present model retains frustration and is therefore not equivalent to the frustration-free Mattis model, its correlated disorder produces an analogous random-gauge structure, as made explicit in the next subsection. The qualifier ``Mattis-like'' is intended to emphasize this similarity. 

In the Kitatani model, this phase would be located along the line $\beta = K_p$, corresponding to the Nishimori line of the EA model and shown as the dotted curve in Fig.~\ref{fig:phase_diagram_Nishimori}. The possible M-SG phase lies in the temperature regime $\beta_{\rm MCP}<\beta<\beta_\text{X}$, between the paramagnetic and ferromagnetic phases.
 Here, $\beta_{\rm MCP}$ represents the multicritical point of the EA model on its Nishimori line, and $\beta_\text{X}$ denotes the ferromagnetic-paramagnetic phase transition point on the Nishimori line of the Kitatani model. 

 The transition point $\beta_\text{X}$ depends on the parameter $a$ in the Kitatani model. The case $a=0$ reduces to the EA model; therefore, $\beta_\text{X}(a=0)=\beta_\text{MCP}$ by definition. In the limit $a\to\infty$, $\beta_\text{X}$ corresponds to the critical disorder strength at which the EA model exhibits a ferromagnetic transition at zero temperature. Since increasing $a$ enhances the ferromagnetic bias of the bond distribution, $\beta_\text{X}(a)$ is expected to be a monotonically increasing function of $a$, with the maximum value achieved in the limit $a\to\infty$. Existing numerical studies of the two-dimensional EA Ising model on a square lattice indicate that the difference between $\beta_\text{MCP}$ and this limiting transition point is not large, namely, $\beta_\text{MCP} \simeq 1.0423$--$1.0495$~\cite{Merz2002, ITO2003262, Picco_2006, Hasenbusch2008, Sasagawa_2020, TNMC2025} and $\beta_\text{X}(\infty)\simeq 1.066$--$1.082$~\cite{Kawashima_1997, WANG200331, Amoruso2004}. This implies that the possible temperature window for the M-SG phase along the Nishimori line is intrinsically narrow. 
 Our primary purpose is to numerically test the proposed existence of this M-SG phase along the dotted curve $\beta=K_p$ in the Kitatani model.

\subsection{Sampling procedure via gauge transformation}
A central challenge in investigating the modified Nishimori model is that the interactions $\bm{\tau}=\{\tau_{ij}\}$ are not i.i.d., but are instead drawn from the correlated distribution given in Eq.~(\ref{eqn:distribution_tau}). Direct sampling from this distribution is not straightforward because it would require evaluating partition-function ratios for each candidate bond configuration. To overcome this difficulty, we reformulate the bond distribution by exploiting the gauge symmetry of the Ising Hamiltonian. Specifically, the distribution in Eq.~(\ref{eqn:distribution_tau}) can be expressed by marginalizing over the configurations of the EA model as follows:
\begin{equation}
    P(\bm{\tau};\gamma,\beta_p) = \sum_{\bm{J},\bm{\sigma}}\prod_{\langle ij\rangle}\delta(\tau_{ij}, J_{ij}\sigma_i\sigma_j)P(\bm{\sigma}|\bm{J},\beta_p)P_{\rm EA}(\bm{J}; \gamma),
\end{equation}
where $\delta$ denotes the Kronecker delta, $P_{\rm EA}(\bm{J};\gamma)$ is the i.i.d. bond distribution of the standard EA model with parameter $\gamma$, and $P(\bm{\sigma}|\bm{J},\beta_p)$ represents the Boltzmann distribution of the gauge spins for a fixed bond configuration $\bm{J}$ at inverse temperature $\beta_p$: 
\begin{equation}
    P(\bm{\sigma}|\bm{J},\beta_p) = \frac{e^{\beta_p\sum_{\langle ij\rangle}J_{ij}\sigma_i\sigma_j}}{Z_{\bm{J}}(\beta_p)}. 
\end{equation}

This representation naturally leads to a hierarchical sampling procedure for the correlated bonds $\bm{\tau}$. First, as in the standard EA model, we generate an auxiliary bond configuration $\bm{J}$ from the distribution $P_{\rm EA}(\bm{J};\gamma)$. Second, given this $\bm{J}$, a gauge configuration $\bm{\sigma}$ is sampled from the conditional Boltzmann distribution $P(\bm\sigma\mid \bm J,\beta_p)$. Finally, by applying the gauge transformation $\tau_{ij}=J_{ij}\sigma_i\sigma_j$, we obtain a bond configuration $\bm{\tau}$ that follows the target distribution $P(\bm{\tau};\gamma,\beta_p)$. Since the transformation by $\bm{\sigma}$ does not change the original frustration pattern of $\bm{J}$, the frustration configuration is determined by $\bm{J}$, while the specific bond realization is dictated by the gauge $\bm{\sigma}$. We refer to this resulting sequence $\bm J\to\bm\sigma\to\bm\tau\to\bm S$ as the hierarchical sampling procedure. 
This hierarchical construction is independent of the specific sampling algorithm. For instance, standard parallel tempering~\cite{Geyer1991, HukushimaNemoto, Marinari1998} can be applied to both $P(\bm\sigma\mid\bm J,\beta_p)$ and $P(\bm S\mid\bm\tau,\beta)$, since the bond configurations are fixed in each conditional simulation. 

\subsection{Numerical sampling scheme}
In this work, we implement the hierarchical sampling procedure using a tensor-network-based (TN) sampler. For both the gauge configurations $\bm{\sigma}$ and the physical spin configurations $\bm{S}$, the method generates independent proposals from approximate conditional distributions and uses importance reweighting to evaluate observables under the corresponding Boltzmann distributions.  

For both types of configurations, the Boltzmann weight is represented as a tensor network. Given an ordering $\pi=(i_1, i_2,\ldots,i_N)$ of the lattice sites, a configuration is generated sequentially using conditional probabilities obtained by contracting the remaining tensor network. The resulting proposal distribution has the autoregressive form 
\begin{equation}
    Q_\chi(\bm{S}) = \prod_{\ell=1}^N Q_\chi\left(S_{i_\ell}\mid S_{i_1},\ldots,S_{i_{\ell-1}}\right). 
\end{equation}
During the contraction, intermediate tensor bonds are truncated to a maximum dimension $\chi$, which controls the tradeoff between the accuracy of the proposal distribution and the computational cost. Because all the conditional probabilities are retained, the proposal probability $Q_\chi(\bm{S})$ of each generated configuration is explicitly available. The same construction is used to generate the gauge configurations $\bm{\sigma}$. 

Recently, hybrid approaches known as tensor network Monte Carlo (TNMC) have been proposed~\cite{TNMHMC2023, TNMC2025}, in which configurations generated approximately using TNs are used to construct a Markov chain. In contrast, we use $Q_\chi$ directly as an independent proposal distribution and account for its deviation from the target distribution through importance reweighting. 
For each target distribution $P(\bm{S})\propto\exp(-\beta H(\bm{S}))$, the TN-based sampler generates a configuration $\bm{S}$ with its corresponding proposal probability $Q_\chi(\bm{S})$. To address the approximation involved in the TN contraction, we evaluate the unnormalized reweighting factor $w$ for each sample,
\begin{equation}
    w(\bm{S}) = \frac{e^{-\beta H(\bm{S})}}{Q_\chi(\bm{S})}. 
\end{equation}
Using these weights, the thermal average of an observable $\mathcal{O}$ for a fixed bond configuration $\bm{\tau}$ is evaluated as a weighted average, 
\begin{equation}
    \langle\mathcal{O}\rangle_{\bm{\tau}} \approx \frac{\sum_{r=1}^{N_\text{spin}}w^{(r)}\mathcal{O}(\bm{S}^{(r)})}{\sum_{r=1}^{N_\text{spin}}w^{(r)}},
\end{equation}
where $N_\text{spin}$ is the number of generated spin configurations and the unknown normalization constant cancels. For two-replica observables, such as the spin-glass overlap, we generate two spin configurations independently for each fixed bond realization $\bm \tau$ and use the product of their importance weights.  

It should be emphasized that the finite bond dimension affects the proposal distribution $Q_\chi(\bm{S})$ rather than replacing the target distribution by a TN approximation. This effect is accounted for by the importance weights, so that the weighted estimator is asymptotically consistent, provided that the proposal has sufficient support. However, an inaccurate proposal can produce significantly nonuniform weights, thus reducing the statistical efficiency and enhancing finite-sample effects. 

We monitor the dispersion of the importance weights using the effective sample size (ESS), defined as 
\begin{equation}
    \text{ESS} = \frac{\left(\sum_{r=1}^{N_\text{spin}}w^{(r)}\right)^2}{\sum_{r=1}^{N_\text{spin}}(w^{(r)})^2}. 
    \label{eqn:ESS}
\end{equation}
An analogous ESS is evaluated for the gauge-configuration sampling. 
In our study, we have confirmed that the ratio $\text{ESS}/N_{\text{spin}}$ remains approximately unity over the investigated parameter range, as shown in Appendix~\ref{sec:appendix}. This indicates that the sampled weights are only weakly dispersed and that the proposals are statistically efficient in the regime studied here. Because each configuration is generated independently rather than via a Markov chain, the method avoids the slow thermalization and autocorrelation issues encountered in conventional MCMC methods. An empirical ESS, however, characterizes only the weights of the generated samples and cannot by itself exclude target-relevant regions that are rarely visited by the proposal or all possible finite-sample effects. We use $\chi=16$ in the following calculations; practical stability checks against the bond dimension and the generation ordering are described in Appendix~\ref{sec:appendix}.

Following this TN-based sampling scheme, we perform hierarchical sampling to incorporate bond correlations. For each frustration pattern $\bm{J}^{(m)}$, we generate $N_{\text{gauge}}$ gauge configurations $\{\bm{\sigma}^{(k)}\}$ and $N_{\text{spin}}$ physical spin configurations $\{\bm{S}^{(k,r)}\}$. For a fixed bond configuration $\bm{\tau}^{(m,k)}$ constructed from the $k$-th gauge configuration and the $m$-th realization of $\bm{J}$, the thermal average of an observable $\mathcal{O}$ is calculated as the reweighted sample average over the generated spin configurations. The overall average, which involves the thermal and disorder averages, is then obtained by  
\begin{equation}
    [\langle \mathcal{O}\rangle ]\approx \frac{1}{N_{\rm frust}}\sum_{m=1}^{N_{\rm frust}}\left(\frac{\sum_{k=1}^{N_{\text{gauge}}}w_{\text{gauge}}^{(m,k)}\langle\mathcal{O}\rangle_{\bm{\tau}^{(m,k)}}}{\sum_{k=1}^{N_\text{gauge}}w_{\text{gauge}}^{(m,k)}}\right),
\end{equation}
where $w_{\text{gauge}}^{(m,k)}$ is the weight for the $k$-th gauge configuration for the $m$-th frustration pattern, and $N_\text{frust}$ is the number of frustration realizations $\bm{J}$.

\subsection{Physical observables}
\label{sec:physical_observables}
To identify the phase transitions and perform finite-size-scaling analyses, we evaluate two distinct order parameters, the magnetization $m$ and the spin-glass order parameter $q$. In the original spin-glass model, these are defined as  
\begin{align}
     m & = \frac{1}{N}\sum_{i=1}^N S_i, \\
       q &= \frac{1}{N}\sum_{i=1}^N S_i^{(1)}S_i^{(2)}, 
\end{align}
where $\bm{S}^{(1)}$ and $\bm{S}^{(2)}$ represent two independent spin configurations under the same interaction realization $\bm{\tau}$. 

In our sampling scheme, where the bonds $\bm{\tau}$ are generated via the gauge transformation $\tau_{ij} = J_{ij}\sigma_i\sigma_j$, the ensemble average is performed over both the frustration patterns $\bm{J}$ and the gauge configurations $\bm{\sigma}$. Under the gauge transformation, the physical spin $S_i$ in the original Hamiltonian is formally replaced by the product $\sigma_i S_i$. Consequently, the expectation value of the magnetization is evaluated as
\begin{equation}
[\langle m\rangle] = \frac{1}{N} \sum_{i=1}^N \langle \sigma_i S_i \rangle_{\bm{J},\bm{\sigma},\bm{S}},
\end{equation}
where the average is taken over $\bm{J}$ at inverse temperature $\gamma$, $\bm{\sigma}$ at $\beta_p$, and $\bm{S}$ at $\beta$. 

Similarly, the second moment of the magnetization required to construct the Binder ratio is expressed as
\begin{equation}
    [\langle m^2\rangle] = \frac{1}{N^2} \sum_{ij}\left[\langle \sigma_i\sigma_j\rangle_{\bm{\sigma}; \beta_p}\langle S_iS_j\rangle_{\bm{S};\beta}\right]_{\bm{J};\gamma}.
\end{equation} 
Generally, the same transformation applies to all even moments: 
\begin{equation}
    [\langle m^{2n}\rangle] = \left[\left\langle \left(
    \frac{1}{N}\sum_i \sigma_i S_i
    \right)^{2n}\right\rangle_{\bm{\sigma},\bm{S}}\right]_{\bm{J}}.
    \label{eqn:moments_gauge}
\end{equation}
This is an exact consequence of the decomposition $\tau_{ij}=J_{ij}\sigma_i\sigma_j$ and the corresponding gauge transformation. Thus, the quantity on the right-hand side is not an alternative definition of the order parameter, but an exact representation of the conventional magnetization moments of the original correlated-disorder model. In particular, the second and fourth moments obtained from Eq.~(\ref{eqn:moments_gauge}) yield the same Binder ratio as the conventional magnetization. Its crossing, therefore, has the usual interpretation as the onset of ferromagnetic order, rather than as a separately defined overlap crossover. 

At the same time, this formulation provides an additional physical interpretation. Because $\bm{\sigma}$ and $\bm{S}$ are equilibrium configurations at inverse temperatures $\beta_p$ and $\beta$, respectively, for the same frustration realization $\bm{J}$, the quantity $N^{-1}\sum_i\sigma_i S_i$ is the overlap between configurations at different temperatures. This representation connects to the concept of temperature chaos discussed in the recent study~\cite{Nishimorietal2025}, without changing its meaning as the conventional ferromagnetic order parameter in the original model. If the system exhibits sensitivity to temperature changes, the correlation between the gauge $\bm{\sigma}$ and the physical spin $\bm{S}$ vanishes, causing the ferromagnetic order parameter to decay. In contrast, the spin-glass order parameter $q$ is invariant under the local gauge transformation because the gauge variables cancel between the two replicas. This invariance has an important consequence. Namely, the disorder-averaged overlap distribution is exactly identical to that of the standard EA model with bond parameter $\gamma$ at the same inverse temperature $\beta$. Indeed, the overlap is unchanged by the gauge transformation, and averaging over the gauge configurations gives $\sum_{\bm{\sigma}}P(\bm{\sigma}\mid\bm{J},\beta_p)=1$. In particular, along the line $\beta=\gamma=K_p$ considered in this study, the overlap distribution coincides with that of the EA model on its Nishimori line, independently of $a$. It is evident that no analogous identity exists for the magnetization $m$.    

For each order parameter $O\in\{m,q\}$, the Binder ratio $g_O$ is a function of both the temperature and the linear dimension, defined as 
\begin{equation}
    \label{eqn:Binder_parameter}
    g_O(\beta,L) = \frac{1}{2}\left(3-\frac{[\langle O^4\rangle]}{[\langle O^2\rangle]^2}\right). 
\end{equation}
The crossing point of $g_O(\beta, L)$ for different system sizes $L$ provides estimates of the corresponding critical inverse temperatures. In particular, a separation between the crossing points of $g_m$ and $g_q$ may signal the existence of an intermediate M-SG phase.  

\section{Numerical results}
\label{sec: Numerical_results}
We present our numerical results for the modified Nishimori model on a square lattice. In our simulations, we focus on the Kitatani model by setting the parameter $a=0.45$. 
The value $a=0.45$ is chosen for practical reasons rather than because it has any special physical significance. At $a=0$, the Kitatani model reduces to the standard EA model and two relevant transition points coincide by construction. Increasing $a$ is expected to widen the possible intermediate phase. However, because the gauge configurations are sampled at $\beta_p=K_p+a$, a larger $a$ also requires TN sampling at lower temperatures, where numerical stability becomes a more significant issue. The chosen value is sufficiently large to make a possible separation numerically accessible while maintaining stable TN sampling used here. 
The simulation parameters for each system size $L$, including the sampling numbers, are summarized in Table~\ref{tab:simulation_setup}. 

\begin{table}[h]
    \centering
    \begin{tabular}{c|lll} 
    \hline 
    $L$  & $N_\text{frust}$ & $N_\text{gauge}$ & $N_\text{spin}$ \\ \hline\hline
    32 & 1000& 1000& 1000\\
    64 & 1000(2000) & 1000 & 1000 \\
    128 & 1000(2000) & 1000 & 2000 \\
    256 & 300(500) & 1000 & 5000 \\ 
    \hline
    \end{tabular}
    \caption{Simulation parameters for each system size $L$. $N_\text{frust}$ denotes the number of frustration realizations $\bm{J}$, $N_\text{gauge}$ the number of gauge configurations $\bm{\sigma}$ per $\bm{J}$, and $N_\text{spin}$ the number of independent spin configurations $\bm{S}$ per $(\bm{J},\bm{\sigma})$. Values in parentheses indicate the increased sample sizes near the transition temperature to enhance statistical accuracy. }
    \label{tab:simulation_setup}
\end{table}

\begin{figure}
    \centering
    \includegraphics[width=\linewidth]{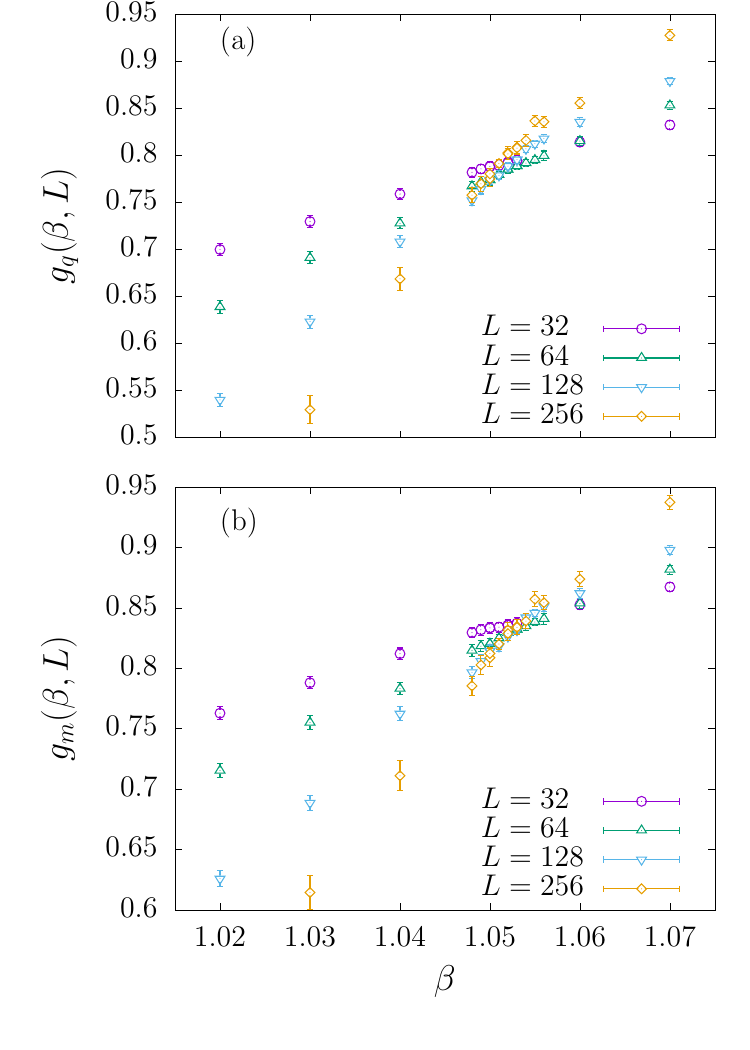}
    \caption{Inverse-temperature dependence of the Binder ratios for (a) the spin-glass order parameter $g_q$ and (b) the ferromagnetic order parameter $g_m$ in the two-dimensional Kitatani model with $a=0.45$. The system sizes examined are $L=32$, $64$, $128$, and $256$. Error bars are smaller than the symbol sizes where they are not visible. }
    \label{fig:Binder_ratio}
\end{figure}

We first examine the Binder ratios $g_q$ and $g_m$ along the line $\beta=K_p$ in the Kitatani model. Figure~\ref{fig:Binder_ratio}(a) shows the inverse-temperature dependence of $g_q(\beta, L)$ for various system sizes. The curves exhibit a crossing near $\beta_c^{q}\simeq 1.050$, indicating the onset of order associated with the SG order parameter. 
In contrast, Fig.~\ref{fig:Binder_ratio}(b) shows that the corresponding crossing region for $g_m$ is shifted slightly toward a higher inverse temperature, near $\beta_c^m\simeq 1.054$. This apparent separation between the crossing regions of $g_q$ and $g_m$ is consistent with the possibility of an intermediate Mattis-like SG phase. 

\begin{figure}
    \centering
    \includegraphics[width=\linewidth]{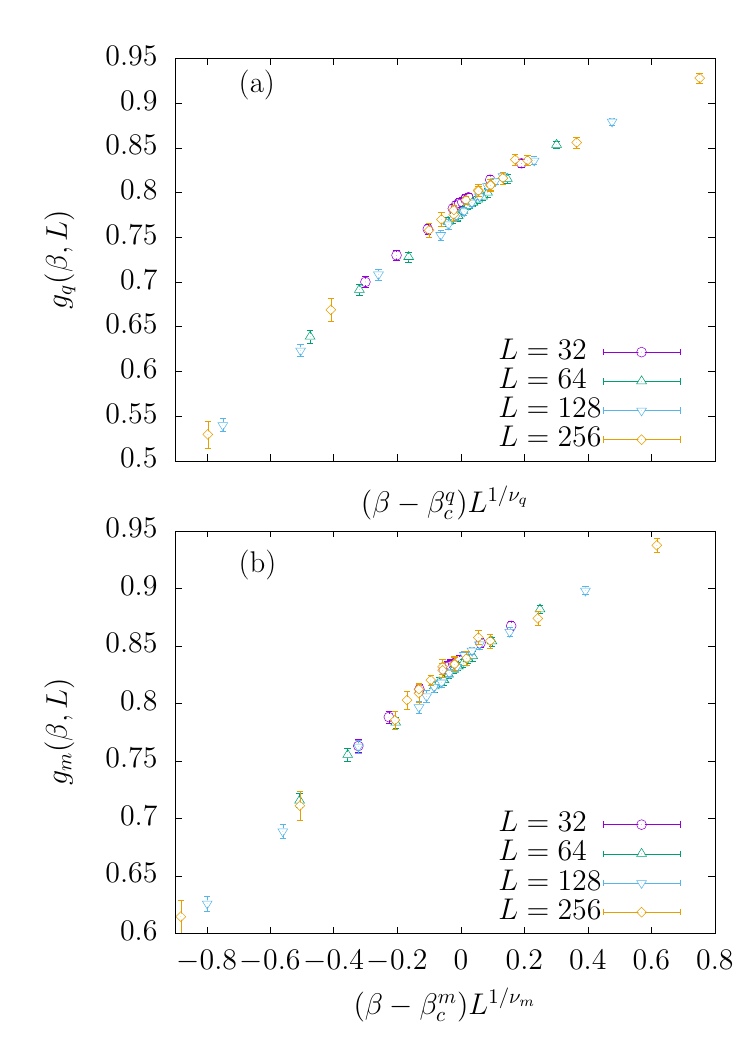}
    \caption{Finite-size-scaling plots for (a) $g_q$ and (b) $g_m$. The horizontal axis represents the scaling variable $(\beta-\beta_c^O)L^{1/\nu_O}$, with the critical point $\beta_c^O$ and $1/\nu_O$ for the order parameter $O$ via Bayesian scaling analysis. The obtained parameters are  $\beta_c^q=1.0506(4)$ and $1/\nu_q=0.654(26)$ for the spin-glass order parameter, and $\beta_c^m=1.0535(3)$ and $1/\nu_m=0.653(25)$ for the ferromagnetic order parameter. }
    \label{fig:Binder_Scaling}
\end{figure}

To quantify this apparent separation and estimate the correlation-length exponents, we perform a finite-size scaling (FSS) analysis using the Bayesian scaling analysis (BSA)~\cite{HaradaBSA2011, HaradaBSA2015} on the Binder ratios. The scaling form used is given by  
\begin{equation}
    g_O(\beta,L) = \tilde{G}_O\left((\beta-\beta_c^O)L^{1/{\nu_O}}\right),
\end{equation}
where $\tilde{G}_O$ is a universal scaling function for the order parameter $O$,  and $\nu_O$ denotes the critical exponent of the correlation length for $O$. This method allows the determination of scaling parameters and their associated uncertainties without explicitly assuming a specific functional form for the scaling function. 

By applying BSA to the spin-glass Binder ratio $g_q$, we obtain the critical inverse temperature $\beta_c^q=1.0506(4)$ and the exponent $1/\nu_q=0.654(26)$, corresponding to $\nu_q\simeq 1.53$. The resulting scaling collapse is shown in Fig.~\ref{fig:Binder_Scaling}(a). Similarly, the BSA for the ferromagnetic Binder ratio $g_m$ yields $\beta_c^m=1.0535(3)$ and $1/\nu_m=0.653(25)$, corresponding to $\nu_m\simeq 1.53$ as shown in Fig.~\ref{fig:Binder_Scaling}(b). The estimated exponents $\nu$ obtained from the two order parameters are consistent with each other and close to the values reported for the multicritical Nishimori point in the standard EA model, where $\nu\simeq 1.5$~\cite{ParisenToldin2009}. This suggests comparable correlation-length scaling at the two estimated transition points, $\beta_c^q$ and $\beta_c^m$, although the agreement of $\nu$ alone does not establish that they are governed by the same underlying fixed point. 

Within the leading finite-size-scaling analysis, the fitted transition-point estimates differ by $\Delta \beta_c=\beta_c^m-\beta_c^q\simeq 0.0029$,  which exceeds their combined quoted statistical uncertainties. However, these uncertainties primarily reflect statistical fluctuations and do not fully account for systematic effects, such as the choice of the scaling window and neglected higher-order corrections to scaling. 
The free boundary conditions employed here may lead to stronger finite-size corrections than periodic boundary conditions. Thus, although the data exhibit good scaling collapse for system sizes up to $L=256$, the apparent separation does not conclusively establish two distinct transitions and may instead reflect observable-dependent finite-size corrections associated with a single transition.   

\begin{figure}
    \centering
    \includegraphics[width=\linewidth]{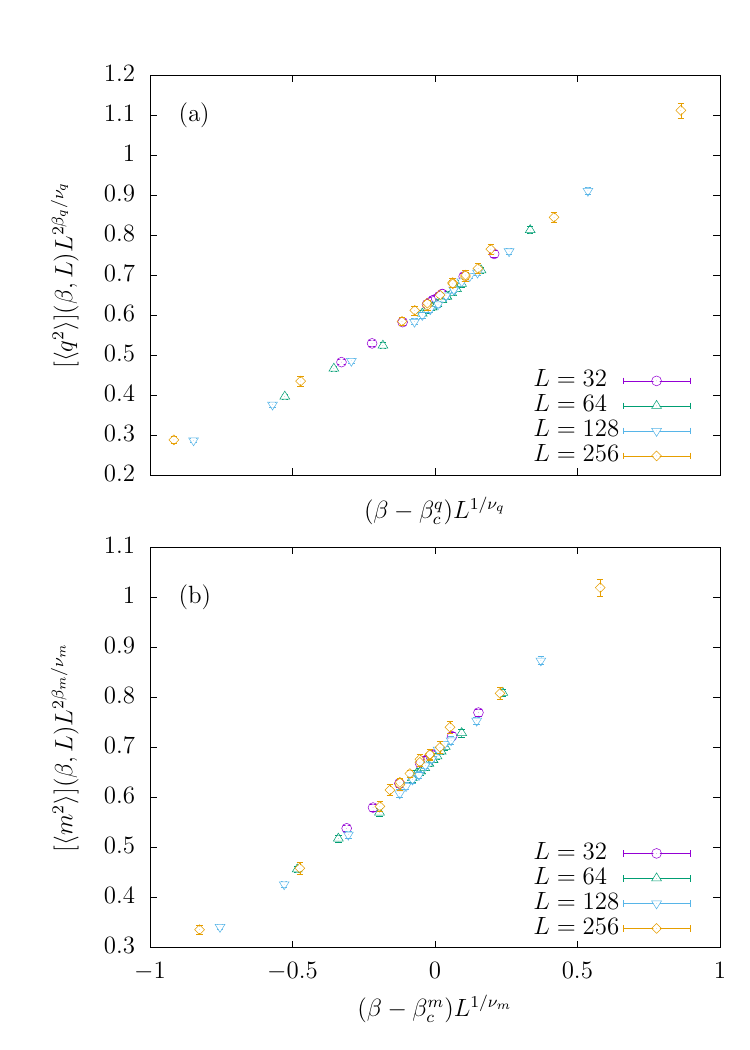}
    \caption{Finite-size-scaling plots of the squared order parameters: (a) spin-glass order $[\langle q^2\rangle]$ and (b) the ferromagnetic order $[\langle m^2\rangle]$. The vertical axis is scaled as $[\langle O^2\rangle]L^{2\beta_O/\nu_O}$ to exhibit the anomalous dimension. The critical inverse temperatures $\beta_c^O$ are fixed to the estimates obtained from the Binder ratio analysis. The estimated scaling parameters are $1/\nu_q=0.684(20)$ and $2\beta_q/\nu_q=0.187(3)$, and $1/\nu_m=0.642(18)$ and $2\beta_m/\nu_m=0.144(2)$. }
    \label{fig:FSS_m2q2}
\end{figure}

To compare the effective scaling behavior associated with the two order parameters, we analyze the FSS of the second moments of the order parameters, $[\langle q^2\rangle]$ and $[\langle m^2\rangle]$. Near the critical point, these quantities follow the scaling form given by 
\begin{equation}
    \left[\langle O^2\rangle\right] = L^{-2\beta_O/\nu_O}\tilde{F}_O\left((\beta-\beta_c^O)L^{1/\nu_O}\right),
\end{equation}
where $\tilde{F}$ denotes the universal scaling function for the order parameter $O$ and $2\beta_O/\nu_O$ is the scaling dimension related to the anomalous dimension $\eta_O$.

Figure~\ref{fig:FSS_m2q2} presents the scaling plots for $[\langle q^2\rangle]$ and $[\langle m^2\rangle]$. To reduce the number of estimated parameters and stabilize the scaling analysis, the critical inverse temperatures $\beta_c^O$ were fixed to the values determined from the Binder ratio analysis. As shown in the figure, we obtain good apparent data collapse within the leading scaling form up to $L=256$. The estimated exponents are $1/\nu_q=0.684(20)$ and $2\beta_q/\nu_q=0.187(3)$ for the spin-glass order parameter and $1/\nu_m=0.642(18)$ and $2\beta_m/\nu_m=0.144(2)$ for the ferromagnetic order parameter.  

\begin{figure}
    \centering
    \includegraphics[width=\linewidth]{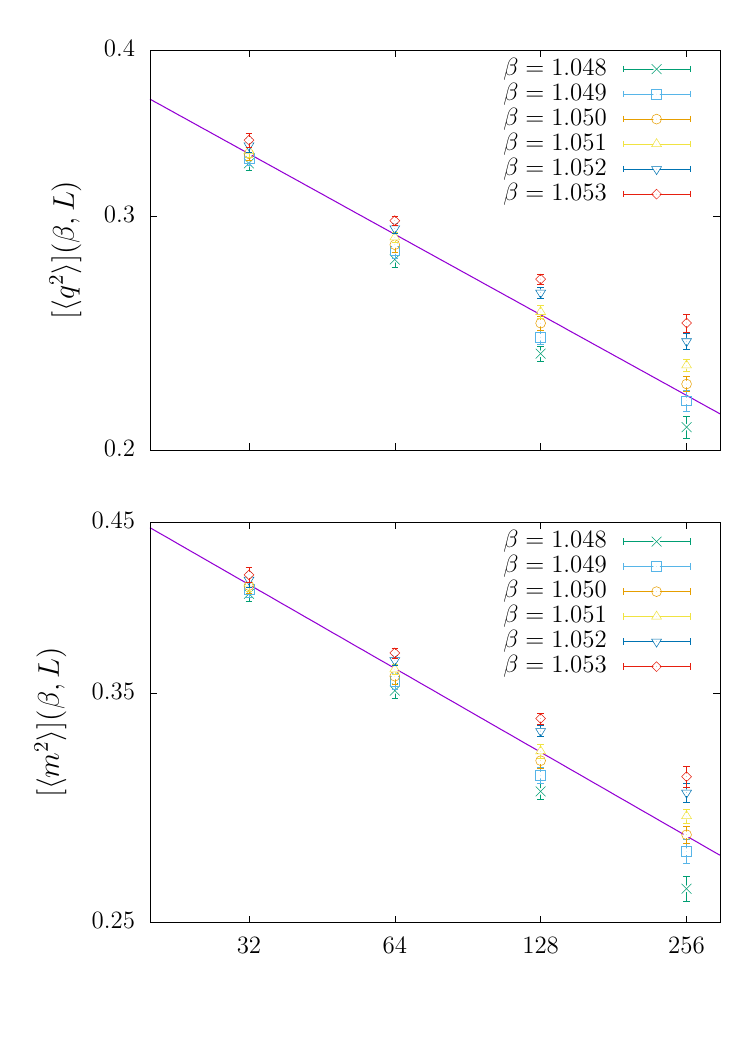}
    \caption{System-size dependence of the second moments of the order parameters, $[\langle q^2\rangle]$ and $[\langle m^2\rangle]$, near the fitted transition-point estimates. The solid lines represent least-squares fits of the data for $L\leq 128$ at $\beta=1.051$ to the power-law form $A_O L^{-x_O}$, where both $A_O$ and $x_O$ are fitting parameters. 
    }
    \label{fig:size-dependences}
\end{figure}

To examine more directly whether the difference between the fitted transition-point estimates corresponds to two distinct transitions, we plot the system-size dependence of $[\langle q^2\rangle]$ and $[\langle m^2\rangle]$ at several fixed values of $\beta$ around the possible transition temperatures in Fig.~\ref{fig:size-dependences}. Under the two-transition scenario, one would expect a range $\beta_c^q<\beta<\beta_c^m$ in which the two order parameters exhibit qualitatively different system-size dependences, corresponding to spin-glass order without ferromagnetic order. However, no such difference is clearly resolved within the accessible system sizes. Instead, the two second moments exhibit similar finite-size trends throughout the narrow interval between the fitted transition-point estimates. This analysis therefore does not provide independent evidence for two distinct transitions. The apparent difference between the fitted transition-point estimates could instead arise from observable-dependent finite-size corrections to scaling in the presence of a single transition. An extremely narrow intermediate Mattis-like SG phase, however, cannot be excluded.  

While the values of $1/\nu$ obtained from the two order parameters are consistent with each other within statistical errors, the estimated values of $2\beta/\nu$ differ, corresponding to anomalous dimensions $\eta_q\simeq 0.187$ and $\eta_m\simeq 0.144$. The former is consistent with the value $\eta\simeq 0.180$ reported at the multicritical Nishimori point of the standard two-dimensional EA model.~\cite{Hasenbusch2008}. This agreement is expected because the gauge invariance of the overlap $q$ makes its disorder-averaged moments identical to those of the standard EA model along the line $\beta=K_p$. For comparison, a smaller value, $\eta\simeq 0.128$, has been reported along the low-temperature paramagnetic-ferromagnetic transition line of the EA model.~\cite{ParisenToldin2009} Our estimate of $\eta_m$ lies between these two reference values.
However, because the transition structure remains unresolved and subleading corrections to scaling have not been systematically incorporated, the present estimates do not allow us to determine the asymptotic critical behavior associated with the ferromagnetic order parameter.  

\section{Conclusion}
\label{sec:conclusion}
In this study, we investigated the phase transitions of a modified Nishimori model on a two-dimensional square lattice, with particular focus on whether an intermediate Mattis-like spin-glass (SG) phase exists along the line $\beta=K_p$. Using a tensor-network (TN)-based hierarchical sampling scheme, we performed high-precision numerical calculations of the SG and ferromagnetic order parameters for system sizes up to $L=256$. 

From a methodological perspective, our approach illustrates the practical advantages of TN-based sampling over traditional MCMC methods. While MCMC often suffers from extremely slow relaxation due to the complex energy landscape of frustrated spin glasses, the TN-based scheme generates independent proposal configurations and corrects their distribution through importance reweighting. Our results demonstrate that this method is effective in the vicinity of the transition region, where conventional methods may encounter severe convergence issues. 

The leading finite-size-scaling analyses of the SG and ferromagnetic Binder ratios yield slightly different transition-point estimates, with a difference exceeding their quoted statistical uncertainties. However, a direct examination of the system-size dependence of the second moments of the two order parameters does not reveal a clear intermediate regime with SG order but without ferromagnetic order. The present data therefore do not allow us to determine whether the fitted estimates correspond to two distinct transitions or arise from observable-dependent finite-size corrections associated with a single transition. Although an extremely narrow intermediate Mattis-like SG phase cannot be excluded, its existence is not established by the present results. Consequently, we cannot draw a definitive conclusion regarding the assumption underlying Kitatani's argument for the absence of reentrant transitions. Although Nishimori's gauge theory provides rigorous constraints on the phase diagram, it does not settle whether a reentrant transition occurs. Our numerical approach therefore provides a complementary and quantitative test of the proposed intermediate Mattis-like SG phase in the correlated-disorder model. Resolving this question will require stronger control of finite-size corrections and possibly larger system sizes.

Regarding the universality class, the critical-exponent estimates obtained from the leading finite-size-scaling analysis differ in part from those at the multicritical point of the standard i.i.d. EA model. While the correlation-length exponent $\nu\simeq 1.5-1.6$ is similar to known values, differences are found in the estimated anomalous dimensions $\eta$. However, the unresolved transition structure and possible corrections to scaling prevent a definitive identification of the universality classes from the present data. From a renormalization-group perspective, the correlated disorder in the modified Nishimori model raises the interesting question of how the spin interactions and the parameters characterizing the disorder distribution should be treated under coarse-graining. Clarifying the resulting renormalization-group flow, together with a more systematic analysis of corrections to scaling, remains an important direction for future work.

\begin{acknowledgments}
The authors would like to express their special thanks to Seiya Miyamoto for numerous and stimulating discussions. We are also grateful to Yoshihiko Nishikawa, Manaka Okuyama, and Masayuki Ohzeki for their helpful comments and suggestions. 
This work was supported by JSPS KAKENHI Grant No.~23H01095 and JST Grant No.~JPMJPF2221.
The computations in this work were partially performed using the facilities of the Supercomputer Center, the Institute for Solid State Physics, the University of Tokyo (ISSPkyodo-SC-2024-Cb-0041, 2025-Ca-0081).
\end{acknowledgments}

\section*{data availability}
The numerical data underlying the results and figures presented in this paper are openly available~\cite{ModifiedIsingSG2026}. 

\bibliography{spin_glass}

@article{Kitatani1992,
author = { Kitatani ,Hidetsugu},
title = {The Verticality of the Ferromagnetic-Spin Glass Phase Boundary of the {$\pm J$} Ising Model in the {$p-T$} Plane},
journal = {Journal of the Physical Society of Japan},
volume = {61},
number = {11},
pages = {4049-4055},
year = {1992},
doi = {10.1143/JPSJ.61.4049},
URL = { 
        https://doi.org/10.1143/JPSJ.61.4049
},
eprint = { 
        https://doi.org/10.1143/JPSJ.61.4049
}
}

@article{Nishimori1981,
    author = {Nishimori, Hidetoshi},
    title = {Internal Energy, Specific Heat and Correlation Function of the Bond-Random Ising Model},
    journal = {Progress of Theoretical Physics},
    volume = {66},
    number = {4},
    pages = {1169-1181},
    year = {1981},
    month = {10},
    issn = {0033-068X},
    doi = {10.1143/PTP.66.1169},
    url = {https://doi.org/10.1143/PTP.66.1169},
    eprint = {https://academic.oup.com/ptp/article-pdf/66/4/1169/5265369/66-4-1169.pdf},
}

@article{Sherrington-Kircpatrick1975,
  title = {Solvable Model of a Spin-Glass},
  author = {Sherrington, David and Kirkpatrick, Scott},
  journal = {Phys. Rev. Lett.},
  volume = {35},
  issue = {26},
  pages = {1792--1796},
  numpages = {0},
  year = {1975},
  month = {Dec},
  publisher = {American Physical Society},
  doi = {10.1103/PhysRevLett.35.1792},
  url = {https://link.aps.org/doi/10.1103/PhysRevLett.35.1792}
}

@article{SFEdwards_1975,
doi = {10.1088/0305-4608/5/5/017},
url = {https://doi.org/10.1088/0305-4608/5/5/017},
year = {1975},
month = {may},
publisher = {},
volume = {5},
number = {5},
pages = {965},
author = {S F Edwards and P W Anderson},
title = {Theory of spin glasses},
journal = {Journal of Physics F: Metal Physics}
}

@book{FarBeyond,
author = {Charbonneau, Patrick and Marinari, Enzo and Mézard, Marc and Parisi, Giorgio and Ricci-Tersenghi, Federico and Sicuro, Gabriele and Zamponi, Francesco},
title = {Spin Glass Theory and Far Beyond},
publisher = {World Scientific},
year = {2023},
doi = {10.1142/13341},
address = {},
edition   = {},
URL = {https://www.worldscientific.com/doi/abs/10.1142/13341},
eprint = {https://www.worldscientific.com/doi/pdf/10.1142/13341}
}

@book{SGbeyond,
author = {Mezard, M and Parisi, G and Virasoro, M},
title = {Spin Glass Theory and Beyond},
publisher = {World Scientific},
year = {1986},
doi = {10.1142/0271},
address = {},
edition   = {},
URL = {https://www.worldscientific.com/doi/abs/10.1142/0271},
eprint = {https://www.worldscientific.com/doi/pdf/10.1142/0271}
}

@book{Diep2013,
author = {Diep, H T},
title = {Frustrated Spin Systems},
publisher = {World Scientific},
year = {2013},
doi = {10.1142/8676},
address = {},
edition   = {2nd},
URL = {https://www.worldscientific.com/doi/abs/10.1142/8676},
eprint = {https://www.worldscientific.com/doi/pdf/10.1142/8676}
}

@article{Nishimori2024,
  title = {Anomalous distribution of magnetization in an Ising spin glass with correlated disorder},
  author = {Nishimori, Hidetoshi},
  journal = {Phys. Rev. E},
  volume = {110},
  issue = {6},
  pages = {064108},
  numpages = {9},
  year = {2024},
  month = {Dec},
  publisher = {American Physical Society},
  doi = {10.1103/PhysRevE.110.064108},
  url = {https://link.aps.org/doi/10.1103/PhysRevE.110.064108}
}

@article{Nishimorietal2025,
  title = {Temperature chaos as a logical consequence of the reentrant transition in spin glasses},
  author = {Nishimori, Hidetoshi and Ohzeki, Masayuki and Okuyama, Manaka},
  journal = {Phys. Rev. E},
  volume = {112},
  issue = {4},
  pages = {044140},
  numpages = {12},
  year = {2025},
  month = {Oct},
  publisher = {American Physical Society},
  doi = {10.1103/qp1w-qcbs},
  url = {https://link.aps.org/doi/10.1103/qp1w-qcbs}
}

@Article{ParisenToldin2009,
author={Parisen Toldin, Francesco
and Pelissetto, Andrea
and Vicari, Ettore},
title={Strong-Disorder Paramagnetic-Ferromagnetic Fixed Point in the Square-Lattice {$\pm J$} Ising Model},
journal={Journal of Statistical Physics},
year={2009},
month={Jun},
day={01},
volume={135},
number={5},
pages={1039-1061},
issn={1572-9613},
doi={10.1007/s10955-009-9705-5},
url={https://doi.org/10.1007/s10955-009-9705-5}
}

@article{Giacomo2011,
  title = {Ferromagnetic-glassy transitions in three-dimensional Ising spin glasses},
  author = {Ceccarelli, Giacomo and Pelissetto, Andrea and Vicari, Ettore},
  journal = {Phys. Rev. B},
  volume = {84},
  issue = {13},
  pages = {134202},
  numpages = {9},
  year = {2011},
  month = {Oct},
  publisher = {American Physical Society},
  doi = {10.1103/PhysRevB.84.134202},
  url = {https://link.aps.org/doi/10.1103/PhysRevB.84.134202}
}

@article{Nobre2001,
  title = {Phase diagram of the two-dimensional {$\pm J$} Ising spin glass},
  author = {Nobre, Fernando D.},
  journal = {Phys. Rev. E},
  volume = {64},
  issue = {4},
  pages = {046108},
  numpages = {6},
  year = {2001},
  month = {Sep},
  publisher = {American Physical Society},
  doi = {10.1103/PhysRevE.64.046108},
  url = {https://link.aps.org/doi/10.1103/PhysRevE.64.046108}
}

@article{Amoruso2004,
  title = {Domain-wall energies and magnetization of the two-dimensional random-bond Ising model},
  author = {Amoruso, C. and Hartmann, A. K.},
  journal = {Phys. Rev. B},
  volume = {70},
  issue = {13},
  pages = {134425},
  numpages = {7},
  year = {2004},
  month = {Oct},
  publisher = {American Physical Society},
  doi = {10.1103/PhysRevB.70.134425},
  url = {https://link.aps.org/doi/10.1103/PhysRevB.70.134425}
}

@article{TNMC2025,
  title = {Tensor network Monte Carlo simulations for the two-dimensional random-bond Ising model},
  author = {Chen, Tao and Guo, Erdong and Zhang, Wanzhou and Zhang, Pan and Deng, Youjin},
  journal = {Phys. Rev. B},
  volume = {111},
  issue = {9},
  pages = {094201},
  numpages = {14},
  year = {2025},
  month = {Mar},
  publisher = {American Physical Society},
  doi = {10.1103/PhysRevB.111.094201},
  url = {https://link.aps.org/doi/10.1103/PhysRevB.111.094201}
}

@Article{TNMHMC2023,
	title={{Collective Monte Carlo updates through tensor network renormalization}},
	author={Miguel Frías-Pérez and Michael Mariën and David Pérez García and Mari Carmen Bañuls and Sofyan Iblisdir},
	journal={SciPost Phys.},
	volume={14},
	pages={123},
	year={2023},
	publisher={SciPost},
	doi={10.21468/SciPostPhys.14.5.123},
	url={https://scipost.org/10.21468/SciPostPhys.14.5.123},
}

@article{HaradaBSA2011,
  title = {Bayesian inference in the scaling analysis of critical phenomena},
  author = {Harada, Kenji},
  journal = {Phys. Rev. E},
  volume = {84},
  issue = {5},
  pages = {056704},
  numpages = {7},
  year = {2011},
  month = {Nov},
  publisher = {American Physical Society},
  doi = {10.1103/PhysRevE.84.056704},
  url = {https://link.aps.org/doi/10.1103/PhysRevE.84.056704}
}

@article{HaradaBSA2015,
  title = {Kernel method for corrections to scaling},
  author = {Harada, Kenji},
  journal = {Phys. Rev. E},
  volume = {92},
  issue = {1},
  pages = {012106},
  numpages = {7},
  year = {2015},
  month = {Jul},
  publisher = {American Physical Society},
  doi = {10.1103/PhysRevE.92.012106},
  url = {https://link.aps.org/doi/10.1103/PhysRevE.92.012106}
}

@article{Wang_2014,
  title = {Topologically invariant tensor renormalization group method for the Edwards-Anderson spin glasses model},
  author = {Wang, Chuang and Qin, Shao-Meng and Zhou, Hai-Jun},
  journal = {Phys. Rev. B},
  volume = {90},
  issue = {17},
  pages = {174201},
  numpages = {9},
  year = {2014},
  month = {Nov},
  publisher = {American Physical Society},
  doi = {10.1103/PhysRevB.90.174201},
  url = {https://link.aps.org/doi/10.1103/PhysRevB.90.174201}
}

@article{Hasenbusch2008,
  title = {Multicritical Nishimori point in the phase diagram of the {$\pm J$} Ising model on a square lattice},
  author = {Hasenbusch, Martin and Toldin, Francesco Parisen and Pelissetto, Andrea and Vicari, Ettore},
  journal = {Phys. Rev. E},
  volume = {77},
  issue = {5},
  pages = {051115},
  numpages = {10},
  year = {2008},
  month = {May},
  publisher = {American Physical Society},
  doi = {10.1103/PhysRevE.77.051115},
  url = {https://link.aps.org/doi/10.1103/PhysRevE.77.051115}
}

@article{ITO2003262,
title = {Nonequilibrium relaxation study on spin glass model},
journal = {Physica A: Statistical Mechanics and its Applications},
volume = {321},
number = {1},
pages = {262-270},
year = {2003},
note = {Statphys-Taiwan-2002: Lattice Models and Complex Systems},
issn = {0378-4371},
doi = {https://doi.org/10.1016/S0378-4371(02)01773-9},
url = {https://www.sciencedirect.com/science/article/pii/S0378437102017739},
author = {Nobuyasu Ito and Yukiyasu Ozeki}
}

@article{Sasagawa_2020,
author = {Sasagawa ,Yoshinori and Ueda ,Hiroshi and Genzor ,Jozef and Gendiar ,Andrej and Nishino ,Tomotoshi},
title = {Entanglement Entropy on the Boundary of the Square-Lattice {$\pm J$} Ising Model},
journal = {Journal of the Physical Society of Japan},
volume = {89},
number = {11},
pages = {114005},
year = {2020},
doi = {10.7566/JPSJ.89.114005},

URL = { 
    
        https://doi.org/10.7566/JPSJ.89.114005
    
    

},
eprint = { 
    
        https://doi.org/10.7566/JPSJ.89.114005
    
    

}
}

@article{Picco_2006,
doi = {10.1088/1742-5468/2006/09/P09006},
url = {https://doi.org/10.1088/1742-5468/2006/09/P09006},
year = {2006},
month = {sep},
publisher = {},
volume = {2006},
number = {09},
pages = {P09006},
author = {Picco, M and Honecker, A and Pujol, P},
title = {Strong disorder fixed points in the two-dimensional random-bond Ising model},
journal = {Journal of Statistical Mechanics: Theory and Experiment}
}

@article{Merz2002,
  title = {Two-dimensional random-bond Ising model, free fermions, and the network model},
  author = {Merz, F. and Chalker, J. T.},
  journal = {Phys. Rev. B},
  volume = {65},
  issue = {5},
  pages = {054425},
  numpages = {18},
  year = {2002},
  month = {Jan},
  publisher = {American Physical Society},
  doi = {10.1103/PhysRevB.65.054425},
  url = {https://link.aps.org/doi/10.1103/PhysRevB.65.054425}
}

@article{Kawashima_1997,
doi = {10.1209/epl/i1997-00318-5},
url = {https://doi.org/10.1209/epl/i1997-00318-5},
year = {1997},
month = {jul},
publisher = {},
volume = {39},
number = {1},
pages = {85},
author = {N. Kawashima and H. Rieger},
title = {Finite-size scaling analysis of exact ground states for {$\pm J$} 
spin glass models in two dimensions},
journal = {Europhysics Letters}
}

@article{WANG200331,
title = {Confinement-Higgs transition in a disordered gauge theory and the accuracy threshold for quantum memory},
journal = {Annals of Physics},
volume = {303},
number = {1},
pages = {31-58},
year = {2003},
issn = {0003-4916},
doi = {https://doi.org/10.1016/S0003-4916(02)00019-2},
url = {https://www.sciencedirect.com/science/article/pii/S0003491602000192},
author = {Chenyang Wang and Jim Harrington and John Preskill}
}

@misc{TNPA,
      title={Tensor-Network Population Annealing}, 
      author={Takumi Oshima and Yuma Ichikawa and Koji Hukushima},
      year={2026},
      eprint={2604.11155},
      archivePrefix={arXiv},
      primaryClass={cond-mat.stat-mech},
      url={https://arxiv.org/abs/2604.11155}, 
      note={To appear in Physical Review E},
}

@article{MATTIS1976421,
title = {Solvable spin systems with random interactions},
journal = {Physics Letters A},
volume = {56},
number = {5},
pages = {421-422},
year = {1976},
issn = {0375-9601},
doi = {https://doi.org/10.1016/0375-9601(76)90396-0},
url = {https://www.sciencedirect.com/science/article/pii/0375960176903960},
author = {D.C. Mattis}
}

@incollection{Marinari1998,
  author    = {Enzo Marinari},
  title     = {Optimized Monte Carlo Methods},
  booktitle = {Advances in Computer Simulation},
  editor    = {J. Kert{\'e}sz and I. Kondor},
  series    = {Lecture Notes in Physics},
  volume    = {501},
  pages     = {50--81},
  publisher = {Springer},
  year      = {1998}
}

@article{HukushimaNemoto,
author = {Hukushima ,Koji and Nemoto ,Koji},
title = {Exchange Monte Carlo Method and  Application to Spin Glass Simulations},
journal = {Journal of the Physical Society of Japan},
volume = {65},
number = {6},
pages = {1604-1608},
year = {1996},
doi = {10.1143/JPSJ.65.1604},

URL = { 
        https://doi.org/10.1143/JPSJ.65.1604
},
eprint = { 
        https://doi.org/10.1143/JPSJ.65.1604
}
}

@inproceedings{Geyer1991,
  author    = {Geyer, Charles J.},
  title     = {Markov Chain Monte Carlo Maximum Likelihood},
  booktitle = {Computing Science and Statistics:
               Proceedings of the 23rd Symposium on the Interface},
  pages     = {156--163},
  year      = {1991}
}

@dataset{ModifiedIsingSG2026,
  author    =   {Takumi Oshima and Yamato Arai and Koji Hukushima},
  title     = {Data and gnuplot scripts for
               "Phase Transitions in a Modified Ising Spin Glass Model:
               A Tensor-Network-based Sampling Approach"},
  year      = {2026},
  publisher = {Zenodo},
  doi       = {10.5281/zenodo.23398138},
  url       = {https://doi.org/10.5281/zenodo.23398138}
}

\appendix

\section{Efficiency and stability of tensor-network-based sampling}
\label{sec:appendix}
To evaluate the statistical efficiency of the tensor-network-based hierarchical sampling, we examine the effective sample size (ESS) ratio defined in Eq.~(\ref{eqn:ESS}). While the concept of ESS is also fundamental in MCMC methods, its interpretation in the present context differs. In MCMC, the ESS commonly quantifies the reduction in the number of effectively independent samples arising from temporal autocorrelation in the Markov chain. In contrast, in our TN-based scheme, which generates configurations independently, the ESS ratio quantifies the dispersion of the reweighting factors (importance weights).  

The ESS ratio evaluates whether the total statistical weight is concentrated on a small subset of potentially ``lucky'' samples. A high ESS ratio indicates that the total weight of the TN-generated samples is not concentrated on a small subset of configurations. This diagnostic is particularly relevant for frustrated systems, where TN contractions can become numerically unstable at low temperatures~\cite{Wang_2014, TNPA}. Such instabilities can lead to an inaccurate proposal and highly nonuniform importance weights. We therefore monitor both the ESS ratio and the stability of the estimated observables with respect to the number of independent samples. These diagnostics can detect domination by exceptionally large weights among the generated samples. They cannot, however, guarantee that all target-relevant regions of configuration space have been explored.  

\begin{figure}
    \centering
    \includegraphics[width=\linewidth]{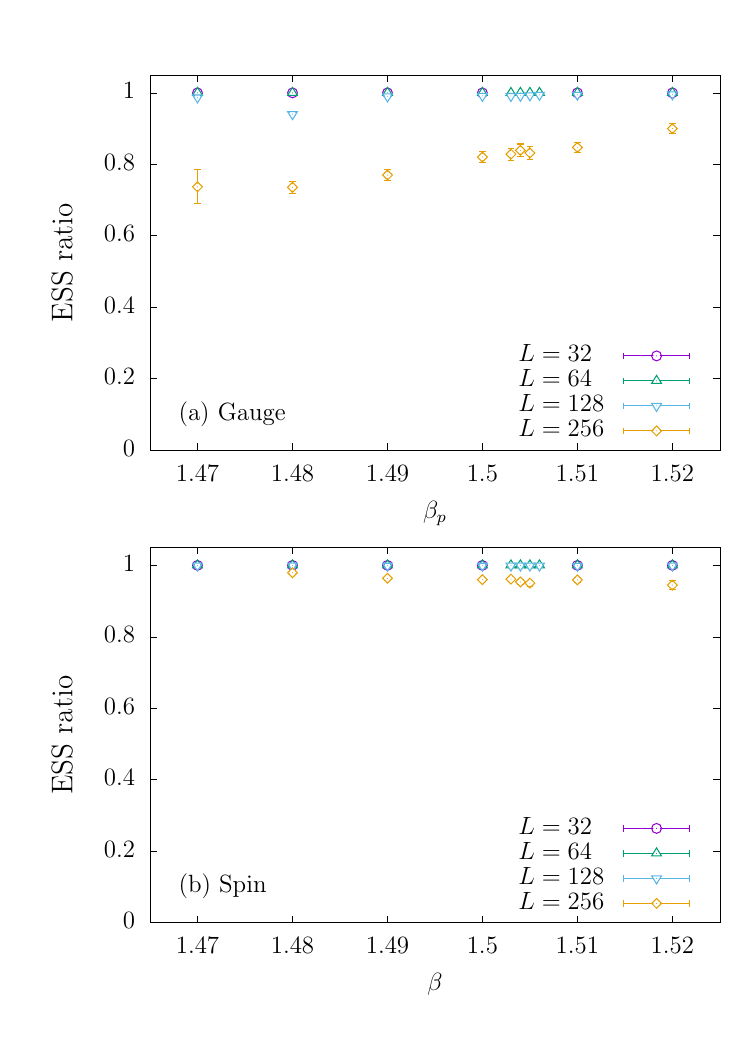}
    \caption{Effective sample size (ESS) per the total number of samples for (a) the gauge-configuration sampling and (b) the spin-configuration sampling. The results are plotted as a function of the corresponding inverse temperature for system sizes $L=32$, $64$, $128$, and $256$ within the temperature range investigated in this study. All results were obtained with bond dimension $\chi=16$. }
    \label{fig:ESS}
\end{figure}

Figure~\ref{fig:ESS} shows the temperature and system-size dependence of the ESS ratio. For gauge-configuration sampling, the ESS ratio exhibits a moderate reduction on the higher-temperature side, particularly for large system sizes, but remains relatively high throughout the investigated range of inverse temperature $\beta$, including the critical region. Furthermore, there is an absence of any significant reduction in the ratio with increasing system size $L$ up to $L=256$. Thus, the TN proposal maintains high statistical efficiency for both gauge- and spin-configuration sampling in the regime studied here. 


The bond dimension $\chi$ controls the accuracy and computational cost of the autoregressive proposal $Q_\chi$ but does not affect the target distribution used to evaluate the reweighting estimator. We use $\chi=16$ in the main calculations. As practical stability checks, we repeated the calculation with $\chi=12$ and with different site-generation orderings, which produce different proposal distributions. The reweighted physical observables reported in the main text agree within their statistical uncertainties in these calculations. These results show no detectable proposal dependence of the reported observables at the statistical resolution of the present calculations, although they do not constitute a rigorous exclusion of all possible rare-event effects.

\end{document}